\documentstyle[preprint,aps,eqsecnum,epsfig]{revtex}
\tighten
\begin{document}
\preprint{PITT-97-1377}
\draft
\title{\bf Gauge Invariant Higgs mass bounds from the Physical Effective Potential}
\author{{\bf D. Boyanovsky, Will Loinaz, R.S. Willey}}   
\address
{Department of Physics and Astronomy,\\ University of
Pittsburgh,\\ Pittsburgh, PA. 15260, U.S.A.}
\date{}
\maketitle
\begin{abstract}
We study a simplified version of the Standard Electroweak Model and introduce the
concept of the physical gauge invariant effective potential in terms of matrix
elements of the Hamiltonian in physical states. This procedure allows an unambiguous
identification of the symmetry breaking order parameter and the resulting effective
potential as the energy in a constrained state. We explicitly compute the physical
effective potential at one loop order and improve it using the RG.  This construction
allows us to extract a reliable, gauge invariant bound on the Higgs mass by unambiguously
obtaining the scale at which new physics  should emerge to preclude vacuum instability.
Comparison is made with popular gauge fixing procedures and an ``error'' estimate is provided between the Landau gauge fixed and the gauge invariant results.  
\end{abstract}
\pacs{12.15.-y;12.15.Ji;11.15.Ex} 
\def\vef{V_{\rm{eff}}}
\def\lef{\lambda_{\rm{eff}}}
\def\svi{s_{\rm VI}}
\def\sew{s_{\rm EW}}
\def\msb{\overline{\rm MS}}
\def\bdm{\begin{equation}}
\def\edm{\end{equation}}
\def\bea{\begin{eqnarray}}
\def\eea{\end{eqnarray}}
\def\NPB#1#2#3{Nucl. Phys. {\bf B#1} #3 (19#2)}
\def\PLB#1#2#3{Phys. Lett. {\bf B#1} #3 (19#2)}
\def\PRD#1#2#3{Phys. Rev. {\bf D#1} #3 (19#2)}
\def\PRL#1#2#3{Phys. Rev. Lett. {\bf#1} #3 (19#2)}
\def\PRT#1#2#3{Phys. Rep. {\bf#1} #3 (19#2)}
\def\MODA#1#2#3{Mod. Phys. Lett. {\bf A#1} #3 (19#2) }
\def\ZPC#1#2#3{Zeit. f\"ur Physik {\bf C#1} #3 (19#2) }


\section{\bf Introduction and Motivation}

Considerable effort has been invested in understanding the
Higgs sector of the Standard Model. Reliable constraints on the Higgs boson
mass are important in determining the energy scales
for collider experiments that probe the electroweak symmetry breaking sector. 
Alternately, should the Higgs be discovered, its properties will help elucidate
high-scale physics. (see\cite{sher} for an early 
review). In the Standard Model the requirement that the conventional
effective potential $\vef(\varphi)$ have its global minimum at the electroweak scale
has been used to obtain a relation between a lower bound on the mass of the 
Higgs boson and the scale at which new physics should appear 
(for recent reviews see\cite{quiros,carena}). That scale is related to the value of 
$\varphi$ (expectation value of the Higgs field) at
which $\vef$ develops a new deeper minimum (which depends on the mass
of the Higgs). Recently Loinaz and Willey\cite{willray} have pointed out
a difficulty with this procedure. When the contributions from the gauge sector of the 
electroweak theory are included in the effective potential, the value of
$\varphi$ which minimizes $\vef$ is {\em gauge dependent}. The gauge dependence of the effective
potential was already recognized by Dolan and Jackiw in their early formulation of effective potentials\cite{dolan}.
Although in some specific gauges the contribution of the gauge sector
may be a perturbative correction to the contributions from the scalar 
plus heavy (top) fermion-Yukawa sector, the gauge dependence implies
that no error estimate can be made and raises questions on the
reliability of such bounds. 
 The conventional $\vef$,
defined in terms of 1PI Green's functions at zero momentum is an off-shell
quantity and in a gauge-fixed formulation is inherently gauge variant and
therefore not uniquely defined. Using the effective potential to study stability or 
metastability implicitly assumes that $\vef$ is associated with the energy of 
(space-time constant) field configurations. In scalar
field theories the effective potential is proven to be the energy of
a constrained state\cite{symanzik,coleman,rivers,suranyi}, but such proof is lacking in gauge theories. 
Since the effective potential as calculated in
gauge-fixed formulations is explicitly gauge
dependent, it cannot generally be identified with the expectation value of the
physical Hamiltonian in a physical state. 
There are, however,   gauge independent quantities that {\em can}
be extracted from the effective potential. Nielsen identities\cite{nielsen} have 
been used to prove that the {\em difference} of the values of the (gauge fixed) 
effective potential evaluated at extrema are gauge 
invariant, as well as the nucleation rate for bubbles in a first order
transition when calculated from the effective potential\cite{metaxas}. 
However providing a lower bound on the Higgs mass requires to estimate
values of the expectation value of the scalar order parameter, which is a {\em gauge variant} quantity
even at the minima of the effective potential. 

Thus it is important to provide 
a formulation in which the effective potential is a gauge invariant
function of a gauge invariant order parameter and can be interpreted as
an energy function. There have been efforts
to formulate a gauge invariant effective action\cite{fradkin} and
consequently an effective potential, but the formalism involved is
formidable and its calculational implementability rather unwieldy. 

Recently a  formulation that allows to obtain a gauge invariant effective
potential as  the expectation value of the Hamiltonian in physical
states has been developed within a different framework\cite{boyan}.  We refer
to this as the Physical Effective Potential (PEP).
In this article we apply the formulation proposed in\cite{boyan} to solve the problem 
of the gauge dependence in a slight variant of the model studied in\cite{willray}. Although this
model is a simplified Abelian version of the Standard Model, it serves to 
demonstrate the utility of the gauge-invariant effective potential 
in providing a gauge-invariant estimate of a vacuum instability scale.

This article is organized as follows: in section II we present the model
to be studied and  the relevant aspects of the gauge invariant formulation 
provided in\cite{boyan} and adapted to include the fermionic sector. In 
this section we define the physical (gauge invariant) observables including the order parameter 
that provides an unambiguous signal for
symmetry breaking. In section III we explicitly construct the one
loop effective potential as the expectation value of the physical Hamiltonian in gauge 
invariant states constrained to give a space-time constant expectation value of the gauge 
invariant order parameter. We also provide the $\msb$ renormalization of this 
effective potential. 
In section IV we compare our results to those obtained from the gauge
fixed formulation in general $R_{\xi,u}$ gauges. 
Section V is devoted to a RG improvement of the gauge invariant effective
potential and to an unambiguous determination of the lower bound on the Higgs mass in this 
model, providing ``error estimates'' for the quantities
obtained in the gauge variant formulation.  In Section VI we present some
brief numerical results.

Section VII summarizes our conclusions and suggests possibles avenues to
extend the gauge invariant construction to non-Abelian gauge theories. 

\section{\bf The Gauge Invariant Description:}

The focus of our study is  the Abelian Higgs model
with an axial coupling of the gauge fields to fermions. The Lagrangian density is
\begin{eqnarray}
 {\cal{L}} & = & -\frac{1}{4}F^{\mu \nu} F_{\mu \nu}+D^{\mu}\phi^{\dagger}
 D_{\mu}\phi-  \mu^2\phi^{\dagger}\phi- \lambda (\phi^{\dagger}\phi)^2+ 
 \bar{\psi}\left[i {D\!\!\!\!/}+y\left(\phi_1+i \gamma^5 \phi_2\right)
\psi \right]  
 \label{higgslag}\\
 D_{\mu}\phi & = & (\partial_{\mu} \phi +i e A_{\mu} \phi) \; ; \; 
{D\!\!\!\!/} = \gamma^{\mu}\left(\partial_{\mu}-ie \gamma^5 \frac{A_{\mu}}{2} \right) \; ; \; \phi= \frac{1}{\sqrt{2}}\left(\phi_1+
i\phi_2\right)
 \label{covder}
\end{eqnarray}

Our goal is to define and explicitly evaluate the {\em gauge invariant}
effective potential, in terms of a gauge invariant order parameter.
The definition of the effective potential that we use is in terms of
the expectation value of the physical Hamiltonian in physical states
constrained to provide a fixed  expectation value of the
physical order parameter. 

 The program for the construction of such an effective
potential requires i) the identification of the physical (gauge
invariant) states of the theory, ii) the identification of an 
 order parameter that is invariant under the {\em local}
gauge transformation but transforms nontrivially under the rigid global symmetry
transformation and thus gives information on spontaneous symmetry breaking
iii) construction of the gauge invariant effective potential as the
expectation value of the Hamiltonian in constrained physical states. Since the concepts 
behind the construction are not part of the standard
lore, we highlight below the most relevant aspects of the formulation,
for more details see\cite{boyan}.

Such a  description is best
achieved within the canonical formulation, which begins with the identification of canonical field variables and constraints. 
These will determine the classical physical phase space and, at the quantum level, the physical Hilbert space.

The canonical momenta conjugate to the scalar and vector fields are given by
\begin{eqnarray}
\Pi^0 & = & 0 \label{pioo} \; ; \;  \Pi^i  = \dot{A}^i+\nabla^i A^0 = -E^i
\label{amom}\\ 
\pi^{\dagger} & = & \dot{\phi}+i e A^0 \phi \; ; \; \pi  =  \dot{\phi}^{\dagger}-i e A^0 \phi^{\dagger} \label{phimom}
\end{eqnarray}

The Hamiltonian is therefore
\begin{eqnarray}
H= && \int d^3x \left\{ \frac{1}{2}\vec{\Pi}\cdot
\vec{\Pi}+\pi^{\dagger}\pi+(\vec{\nabla}\phi- ie \vec{A}\phi)\cdot
(\vec{\nabla}\phi^{\dagger}+ie \vec{A}\phi^{\dagger})+\frac{1}{2} (\vec{\nabla}
\times \vec{A})^2 + \right. \nonumber \\ 
&&\left. \mu \phi^{\dagger}\phi+ \lambda
(\phi^{\dagger}\phi)^2 + A_0\left[\vec{\nabla}\cdot \vec{\Pi} - ie (\pi
\phi-\pi^{\dagger}\phi^{\dagger})+i\frac{e}{2}\psi^{\dagger}\gamma^5
\psi\right]+ \right. \nonumber \\
&& \left. \psi^{\dagger}\left[-i\vec{\alpha}\cdot\left(
\vec{\nabla}-i\frac{e}{2}\gamma^5 \vec{A}\right)+\beta y \left(\phi_1+i
\gamma^5 \phi_2\right)\right]\psi \right\} \label{hamiltonian}
\end{eqnarray}

where $\vec{\alpha} \; ; \beta$ are the Dirac matrices. 

There are two main methods for quantizing gauge theories, the first one
originally due to Dirac\cite{dirac}(see also\cite{jackiw2}) begins
by identifying the first class (mutually commuting) constraints and 
projects the physical states by requiring that these are simultaneously
annihilated by all first class constraints. Physical operators commute with all of the first class 
constraints. The second, and most used method, ``fixes a gauge'', converting the set of 
first class constraints
into second class constraints (non-commuting) and introducing the
Dirac brackets.  This is the popular
method of dealing with the constraints and leads to the usual gauge-fixed path integral 
representation\cite{slavnov} in terms of Faddeev-Popov determinants and ghosts. Although 
this second method is the most popular as
it is easily translated into a path integral language, it has the drawback
that the physical quantities are more difficult to extract, and though
S-matrix elements are gauge invariant, off-shell quantities generally are not. 

In order to avoid ambiguities and to define a physical order parameter and
effective potential (an off-shell quantity) we choose to use the first method.  

In Dirac's method of quantization\cite{dirac,jackiw2} there are two first class constraints which are:
\begin{equation} 
\Pi^0= \frac{\delta {\cal{L}}}{\delta A^0} = 0 \label{pi0const}
\end{equation}
and Gauss' law:
\begin{eqnarray}
{\cal{G}}(\vec{x},t) & = & \vec{\nabla}\cdot \vec{\Pi} - \rho=0 \label{gausslaw} \\ 
\rho & = & \left[ie (\pi
\phi-\pi^{\dagger}\phi^{\dagger})-i\frac{e}{2}\psi^{\dagger}\gamma^5
\psi\right] \label{rho}
\end{eqnarray}
with $\rho$ being the matter (complex scalar and fermionic) field charge density.

Gauss' law can be seen to be a constraint in two ways: either because it cannot be 
obtained as a Hamiltonian equation of motion, or because in Dirac's formalism, it is the 
secondary (first class) constraint obtained by requiring that the primary constraint 
(\ref{pi0const}) remain constant in
time. Quantization is now achieved by imposing the canonical equal-time
commutation relations
\begin{equation}
\left[\Pi^0(\vec{x},t),A^0(\vec{y},t)\right]  =  -i \delta(\vec{x}-\vec{y})
\; \; ; \; \; \left[\Pi^i(\vec{x},t),A^j(\vec{y},t)\right]  =  -i
\delta^{ij} \delta(\vec{x}-\vec{y}) \label{acomrel}\\
\end{equation}
along with the usual canonical commutators for the scalar field and its canonical momentum 
and anticommutators for the fermionic fields.

In Dirac's formulation, the projection onto the gauge invariant subspace of the full Hilbert 
space is achieved by imposing the first class constraints onto the states.
 Physical operators are those that commute with the first class
constraints. Since $\Pi^0(\vec{x},t)$ and ${\cal{G}}(\vec{x},t)$ are generators of 
local gauge transformations, operators that commute with the first class 
constraints are gauge invariant\cite{boyan}.

Following the steps of reference\cite{boyan} we find that 
 the fields and canonical conjugate momenta
\begin{eqnarray}
\Phi(\vec{x}) & = & \phi(\vec{x})\exp\left[ie \int d^3 y \vec{A}(\vec{y})\cdot
\vec{\nabla}_y G(\vec{y}-\vec{x})\right]= \frac{1}{\sqrt{2}} \left(\Phi_1+ i \Phi_2\right) \label{gauginvphi} \\
\Pi(\vec{x}) & = & \pi(\vec{x}) \exp\left[-ie \int d^3 y \vec{A}(\vec{y})\cdot
\vec{\nabla}_y G(\vec{y}-\vec{x})\right] \label{canophi} \\
\Psi(\vec x) & = & \exp\left[-i\frac{e}{2}\gamma^5 \int d^3 y
\vec{A}(\vec{y})\cdot \vec{\nabla}_y G(\vec{y}-\vec{x})\right]
\psi(\vec x) \label{gauginvpsi}
\end{eqnarray}
with $G(\vec{y}-\vec{x})$ the Coulomb Green's function that satisfies
\begin{equation}
\nabla^2 G(\vec{y}-\vec{x})=\delta^3 (\vec{y}-\vec{x}) \label{coulombgf}
\end{equation}
are invariant under the gauge transformations\cite{boyan}.
Furthermore writing the gauge field into transverse and longitudinal
components as follows
\begin{eqnarray}
&& \vec{A}(\vec{x}) = \vec{A}_L(\vec{x})+\vec{A}_T(\vec{x}) \label{fieldsplit}
\\ && \vec{\nabla} \times \vec{A}_L(\vec{x}) = 0 \; \; ; \; \; 
\vec{\nabla} \cdot \vec{A}_T(\vec{x}) = 0 \label{trans}
\end{eqnarray}
it is clear that
 the ``transverse component'' $\vec{A}_T(\vec{x})$ is also  a gauge invariant operator. 
The fields $\vec{A}_T \; ; \Phi \; ;  \Psi$  and their canonical
momenta are {\em gauge invariant}
as they commute with the constraints. 
The momentum canonical to $\vec{A}\; ; \; \vec{\Pi}$ is written in terms of
``longitudinal'' and ``transverse'' components
\begin{equation}
\vec{\Pi}(\vec{x}) = \vec{\Pi}_L(\vec{x}) +\vec{\Pi}_T(\vec{x}) \label{PiA}.
\end{equation} 
Both components are gauge invariant.

In the physical subspace of gauge invariant wave-functionals, matrix elements
of $\vec{\nabla}\cdot \vec{\Pi}$ can be replaced by matrix elements of the
charge density $\rho$, since matrix elements of Gauss' law between 
these states vanish. Therefore in all matrix elements between gauge
invariant states (or functionals) one can replace
\begin{equation}
\vec{\Pi}_L(\vec{x}) \rightarrow  \rho
\label{coulomb}
\end{equation}
with the charge density (a gauge invariant operator) written in terms of
the gauge invariant fields as
\begin{equation}
\rho=\left[ ie
\left(\Phi(\vec{y})  \Pi(\vec{y})  - \Phi^{\dagger}(\vec{y}) \Pi^{\dagger}(\vec{y}) \right)-i\frac{e}{2}\Psi^{\dagger}(\vec{y}) \gamma^5
\Psi(\vec{y}) \right]
\label{rhogaugeinv}
\end{equation}
This procedure is tantamount to solving the constraints in the physical
space\cite{jackiw2}. 

Finally in the gauge invariant subspace the Hamiltonian becomes
\begin{eqnarray}
H= && \int d^3x \left\{ \frac{1}{2}\vec{\Pi}_T\cdot
\vec{\Pi}_T+\Pi^{\dagger}\Pi+(\vec{\nabla}\Phi- ie \vec{A}_T \Phi)\cdot
(\vec{\nabla}\Phi^{\dagger}+ ie \vec{A}_T \Phi^{\dagger})+\frac{1}{2}
(\vec{\nabla} \times \vec{A}_T)^2 + \right. \nonumber \\
 && \left. \mu^2 \Phi^{\dagger} \Phi+ \lambda
(\Phi^{\dagger}\Phi)^2 +\Psi^{\dagger}\left[-i\vec{\alpha}\cdot\left(
\vec{\nabla}-i\frac{e}{2}\gamma^5 \vec{A}_T\right)+\beta y \left(\Phi_1+i
\gamma^5 \Phi_2\right)\right]\Psi \right\} + \nonumber \\
&&\frac{1}{2}\int d^3y \int d^3x
\rho(\vec{x})G(\vec{x}-\vec{y})\rho(y)
\label{gauginham}
\end{eqnarray}
Clearly the Hamiltonian is gauge invariant, and it manifestly has the global $U(1)$ (chiral) symmetry under which 
\begin{equation}
\Phi(\vec x) \rightarrow \Phi(\vec x) e^{i \varphi} \; \; ; \; \; 
\Psi(\vec x) \rightarrow e^{-i\gamma^5 \frac{\varphi}{2}}\Psi(\vec x)
\label{globaltransf}
\end{equation}
with $\varphi$ a constant real phase,   
$\Pi$ transforms with the opposite phase and
$\vec{A}_T$ is invariant.

The Hamiltonian written in terms of gauge invariant field operators
(\ref{gauginham}) is reminiscent of the Coulomb gauge Hamiltonian, but we
emphasize that we have not imposed any gauge fixing condition. The formulation
is fully gauge invariant, written in terms of operators that commute with the 
generators of gauge transformations and states that are invariant under these 
transformations. The similarity to the Coulomb gauge Hamiltonian is
a consequence of the fact that in this Abelian theory, Coulomb gauge displays 
explicitly the {\em physical} degrees of freedom. 

There is a definite advantage in this gauge invariant formulation: the
(composite) field $\Phi(\vec{x})$ is a candidate for a {\em locally gauge
invariant order parameter}. The point to stress is the following: this operator 
is {\em invariant} under local gauge transformations 
but it transforms as a charged operator under the {\em global} gauge
transformations generated by $Q=\int d^3x \rho(\vec{x})$, that is
\begin{equation}
e^{i\alpha Q}\Phi(\vec{x}) e^{-i\alpha Q}= e^{ie\alpha} \Phi(\vec{x}).
\end{equation}

Because the gauge constraints annihilate the physical states and these
constraints are the generators of local gauge transformations\cite{boyan,jackiw2},
 these states are invariant under the local gauge transformations and any operator that 
{\em is not} invariant under these local transformations {\em must} have zero
expectation value. The {\em local} gauge symmetry cannot be spontaneously
broken; this result is widely known in lattice gauge theory as Elitzur's
theorem\cite{elitzur}. However, the {\em global} symmetry generated by the
charge $Q$ {\em can} be spontaneously broken and the expectation value of a
charged field signals this breakdown.

From this discussion we clearly see that a trustworthy order parameter must
be invariant under the local gauge transformations, thus commuting with the
gauge constraints, but must transform non-trivially under the global gauge
transformation generated by the charge.  The field $\Phi$ fulfills these
criteria and is the natural candidate for an order parameter.

At this stage, having recognized the physical states one could prefer
to pass to a path integral representation of the vacuum-in to vacuum-out 
transition amplitude. This can now be done unambiguously by carrying out
the usual procedure in terms of phase space path integrals with the
gauge invariant measure ${\cal{D}}\vec{\Pi}_T {\cal{D}}{\vec A}_T {\cal{D}}\Phi {\cal{D}}\Phi^{\dagger} {\cal{D}}\Pi {\cal{D}}\Pi^{\dagger}\cdots $. There is no need for ``gauge fixing''. 
In the resulting action (of the form $p\dot{q}-H$), the instantaneous
Coulomb interaction can be re-written by introducing an auxiliary field,
and the integral over the canonical momenta can be carried out explicitly
leading to a Lagrangian form. The resulting Lagrangian  leads to
Feynman rules that are very similar to those in Coulomb gauge and allow
the perturbative calculation of wave function renormalization constants needed below. 
The gauge invariant effective potential can also be computed  in this
path integral representation, but we prefer to provide its explicit 
construction from the Hamiltonian as such construction displays more
clearly the identification of the effective potential as the energy
of a constrained state.

\section{\bf The Effective Potential}

We are now in position to define the gauge invariant effective
potential. Consider the class of
gauge invariant states $|\Upsilon,\chi \rangle$ characterized by
the condition that
the expectation value of the gauge invariant order parameter $\Phi(\vec{x})$ in
this state is nonzero and space-time constant
\begin{equation}
\frac{\langle \Upsilon, \chi| \Phi(\vec{x})| \Upsilon,\chi\rangle}{\langle \Upsilon, \chi
| \Upsilon,\chi \rangle}= \frac{1}{\sqrt{2}}(\chi_1+i\chi_2) \equiv \chi  \label{expval}
\end{equation}
In this notation $\Upsilon$ indexes the states within the set characterized by
eq. \ref{expval}.
The effective potential is defined as the minimum of the expectation 
value of the Hamiltonian density over this class of 
constrained states \cite{symanzik,coleman,rivers}, namely
\begin{equation}
\vef(\chi) = \frac{1}{\Omega} \min_{\Upsilon}\left\{ \frac{\langle \Upsilon,\chi| H
| \Upsilon,\chi\rangle}{\langle \Upsilon, \chi |  \Upsilon,\chi \rangle} \right\}
\label{gauginveff}
\end{equation}
with $H$ being the gauge invariant Hamiltonian given by equation
(\ref{gauginham}) and $\Omega$ the spatial volume\cite{suranyi}. 
By construction in terms of gauge invariant states (or functionals) and
the gauge invariant Hamiltonian, this effective potential is {\em gauge invariant}. 
The minima of this effective potential are obtained from $\vef(\chi)$ by further
minimizing with respect to $\chi_{1,2}$.

It is convenient to separate the expectation value of $\Phi$ as
\begin{eqnarray}
\Phi(\vec{x}) & = & \frac{1}{\sqrt{2}}(\chi_1+i\chi_2) + \frac{1}{\sqrt{2}}
 (\eta_1(\vec{x})+i\eta_2(\vec{x})) \label{separ}\\
\Pi(\vec x) & = & \frac{1}{\sqrt{2}}(\Pi_1(\vec x) -i \Pi_2(\vec x))
\label{pisepar}
\end{eqnarray}
The one-loop correction (formally of ${\cal{O}}(\hbar)$) to the effective
potential is obtained by keeping the {\em linear and quadratic} terms in
$\eta_{1,2}$ in the Hamiltonian, however the linear terms will not contribute 
to the effective potential because their contribution vanishes
upon taking the expectation value in the state $|\Upsilon,\chi\rangle$. Thus keeping
only the quadratic terms in $\eta_{1,2}$ we obtain

\begin{eqnarray}
H_q & = & \Omega \left( \frac{\mu^2}{2}|\chi|^2 + 
\frac{\lambda}{4} |\chi|^4 \right)+ H_q^B +H^F \nonumber \\
H_q^B & = & \frac{1}{2}
\int d^3x \left\{ 
\vec{\Pi}^2_T+  (\vec{\nabla} \times \vec{A})^2 +
 e^2|\chi|^2  \vec{A}^2_T +
\Pi^2_1+\Pi^2_2 +
(\vec{\nabla}\eta_1)^2+ (\vec{\nabla}\eta_2)^2 + \right. \nonumber \\
&& \left. \eta^2_1(\mu^2+3 \lambda |\chi|^2)+\eta^2_2(\mu^2+\lambda |\chi|^2) \right\}- \frac{e^2}{2} |\chi|^2 \int d^3x \int d^3y 
\Pi_2(\vec x)  G(\vec x - \vec y)\Pi_2(\vec y) \nonumber \\
H^F & = & \int d^3 x \left\{ \Psi^{\dagger}\left[ -i\vec{\alpha}\cdot \nabla + y |\chi|\beta \right] \Psi \right\} 
\label{quadham}
\end{eqnarray}
where we have performed a {\em rigid} chiral phase rotation under which the Hamiltonian is invariant. 
The transverse components $\vec{A}_T$ describe a field with mass 
$m_T^2= e^2|\chi|^2$ and only two polarizations. The fermionic part is 
recognized as a free Dirac fermion with mass $M_F= y|\chi|$. The fermionic 
Hamiltonian can be diagonalized in terms of 
particle and antiparticle creation and annihilation operators of the usual form, 
and upon using their anticommutation relation we obtain
\begin{eqnarray}
H^F & = & \sum_{k} \sum_{a=1,2}\left\{\omega^F(k) \left[ b^{\dagger}_{k,a}b_{k,a}+d^{\dagger}_{k,a}d_{k,a}\right]\right\}-
2 \sum_k \omega^F(k) \label{fermionpart} \\
\omega^F(k) & = & \sqrt{k^2+M^2_F[|\chi|]} \; \; ; \; \; M_F[\chi]=y|\chi|
\label{fermionfreq}
\end{eqnarray}

 Following\cite{boyan} we write the bosonic Hamiltonian ($H_q^B$)
  in terms of the spatial Fourier
transform of the fields and their canonical momenta in terms of which the 
quadratic part of the Hamiltonian finally becomes

\begin{eqnarray}
H_q & = & \Omega V_{cl}(|\chi|)+ \frac{1}{2}\sum_{k}\left\{\vec{\Pi}_T(k) \cdot
\vec{\Pi}_T(-k)+ \omega^2_T(k) \vec{A}_T(k) \cdot \vec{A}_T(-k) +
\right. \nonumber \\ & & \left. \Pi_1(k) \Pi_1(-k)+ \omega^2_H(k) \eta_1(k)
\eta_1(-k) + \Pi_2(k) \Pi_2(-k) \frac{\omega^2_T(k)}{k^2}+ \eta_2(k) \eta_2(-k)
\omega^2_g (k) \right\} \label{quadhamk}
\end{eqnarray}
where the frequencies are given in terms of the effective masses as
\begin{eqnarray}
\omega^2_T(k) = k^2+M^2_T[\chi] &\;\;;\;\;& M^2_T[\chi] = e^2|\chi|^2
  \label{transfreq} \\
\omega^2_H(k) = k^2+M^2_H[\chi]&\;\;;\;\;& M^2_H[\chi] =\mu^2 +
 3 \lambda |\chi|^2
  \label{realfreq} \\
\omega^2_g(k) = k^2+M^2_g[\chi]&\;\;;\;\;& M^2_g[\chi] = \mu^2 +\lambda |\chi|^2 .
  \label{imfreq}
\end{eqnarray}
The last two terms can be brought to a canonical form by a Bogoliubov transformation. 
Define the new canonical coordinate $Q$ and conjugate momentum $P$ as
\begin{equation}
\Pi_2(k)  =  \frac{k}{\omega_T(k)} P(k) \; \; ; \; \; 
 \eta_2(k)  = 
\frac{\omega_T(k)}{k} Q(k) \label{newmomcoord}
\end{equation}
in terms of which the last term of the Hamiltonian (\ref{quadhamk}) becomes a canonical quadratic form with the {\em plasma} frequency
\begin{equation}
\omega^2_p(k) = \omega^2_g(k) \frac{\omega^2_T(k)}{k^2}=
 [k^2+M^2_g[\chi]][k^2+M^2_T[\chi]]/k^2 \label{plasmafreq}
\end{equation}

There are four physical degrees of freedom. The modes with frequency
$\omega_T(k)$ are the two transverse degrees of freedom, the mode with
frequency $\omega_H(k)$ is identified with the Higgs mode. In absence of
electromagnetic interactions ($e=0$) the mode with frequency $\omega_p(k)$
represents the Goldstone mode whereas {\em in equilibrium}, namely at the
minimum of the tree level potential, when $|\chi| = \sqrt{-\mu^2/\lambda}$, it represents the 
plasma mode which is identified as the screened Coulomb interaction, and the
transverse and plasma modes all share the same mass. A detailed discussion of the 
dispersion relation of the bosonic excitations has been provided 
in\cite{boyan}.

The quadratic Hamiltonian is now diagonalized in terms of creation and
destruction operators for the quanta of each harmonic oscillator. The ground
state is the vacuum for each oscillator and is the state of lowest
energy compatible with the constraint (\ref{expval}).
 Therefore the one loop (${\cal{O}}(\hbar)$) contribution to the
effective potential is obtained from the zero point energy of the bosonic
oscillators and the (negative) contribution from the ``Dirac sea''
given by the last term in (\ref{fermionpart}). Therefore accounting for the two polarizations 
of the transverse components we find:
\begin{equation}
\vef(|\chi|) = V_{cl}(|\chi|)+ \frac{1}{2} \int \frac{d^3k}{(2\pi^3)}[2
\omega_T(k)+ \omega_H(k)+\omega_p(k)-4\omega^F(k)] \label{gauginveffpot}
\end{equation}

The normalized state that satisfies (\ref{expval}) and gives the
minimum expectation value of the Hamiltonian, thus determining effective
potential via (\ref{gauginveff}) is given by
\begin{equation}
|\chi \rangle = |\chi\rangle_{T_1}\otimes|\chi\rangle_{T_2}\otimes
|\chi\rangle_{H}\otimes|\chi\rangle_{p}\otimes|\chi\rangle_{F}
\end{equation}
i.e. a tensor product of the harmonic oscillator ground states for
the two polarizations ( $T_{1,2}$ ), the Higgs mode (H), the ``plasma''
mode (p) and the fermionic Fock state (F) (the ``Dirac sea''). The functional representation 
for the bosonic Fock states is simply a  Gaussian wave-functional. 
This state {\em by construction} is gauge invariant. 

The k-integrals in the final form of the gauge invariant effective potential
(\ref{gauginveffpot}) are carried out in dimensional regularization 
by the replacements:  $\int d^3k/(2\pi)^3 \rightarrow \int d^{d-1}k / (2\pi)^{d-1} ; g_{\mu}^{\mu}=d$, and $2 \rightarrow d-2$ transverse modes, $d=4-\epsilon$ (see
the appendix) and we obtain the following (unrenormalized expression) one
loop effective potential
\begin{eqnarray}
\vef(\chi)  =  V_{cl}(\chi)& + & \frac{1}{(4\pi)^2}\left\{
-\Delta_{\epsilon}\left[ \frac{M^4_T[\chi]}{2}+ \frac{M^4_H[\chi]}{4}+
\frac{1}{4} (M^2_T[\chi]-M^2_g[\chi])^2-M^4_F[\chi] \right]+ \right. 
\nonumber \\
 &  & \left. \frac{M^4_T[\chi]}{2}\left(\ln\left[\frac{M^2_T[\chi]}{\kappa^2}\right]-\frac{1}{2} \right)+ \frac{M^4_H[\chi]}{4}\left(\ln\left[\frac{M^2_H[\chi]}{\kappa^2}\right]-\frac{3}{2} \right)+ \right. \nonumber \\
& & \left. \frac{1}{4} \left(M^2_T[\chi]-M^2_g[\chi][\chi]\right)^2
\left(\ln\left[\frac{(M_T[\chi]+M_g[\chi])^2}{\kappa^2}\right]-\frac{3}{2} \right) \right.\nonumber \\
& &  \left. 
-\frac{1}{2}M_T[\chi]M_g[\chi](M_T[\chi]-M_g[\chi])^2-  M^4_F[\chi]\left(\ln\left[\frac{M^2_F[\chi]}{\kappa^2}\right]-\frac{3}{2} \right)
\right\} \label{veffinv}
\end{eqnarray}
with
\begin{equation}
\Delta_{\epsilon} = \frac{2}{4-d}-\gamma + \ln[4\pi]\label{poleterm}
\end{equation}
is the standard $\msb$ UV subtraction in d-dimensions,
$\kappa$ is the renormalization scale and $\gamma$ is the Euler-Mascheroni
constant. 

\subsection{$\msb$ Renormalization}

The theory is renormalized by coupling, mass and wavefunction renormalizations.

\begin{equation}
e_b  =   Z_e e_R \; ; \; \lambda_b = Z_{\lambda}\lambda_R \; ; \; 
y_b = Z_y y_R \; ; \; \mu^2_b = Z_{\mu^2} \mu^2_R \label{renorcoups}
\end{equation}

\begin{equation}
A^{\mu}_b = Z^{\frac{1}{2}}_A A_R \; ; \; \Phi_b = Z^{\frac{1}{2}}_{\Phi}
\Phi_R \; ; \; \Psi_b = Z^{\frac{1}{2}}_{\Psi}\Psi_R
\label{wavefuncs}
\end{equation}

The gauge invariant quantization ensures that all the Ward identities
associated with the gauge symmetry are now trivially satisfied. Here
we work in the $\msb$ renormalization scheme and write
for the various couplings and wavefunction renormalizations $Z=1+\delta Z$,
expanding $\delta Z$ in powers of the couplings and absorbing only the
$\msb$ divergences proportional to $\Delta_{\epsilon}$. 
There are two equivalent procedures to renormalize the effective 
potential:
\begin{enumerate}
\item
The terms proportional to $\Delta_{\epsilon}$ in (\ref{veffinv}) are absorbed in a 
partial renormalization of couplings and $\mu^2$.  This renormalization,
however, does not render finite the scattering amplitudes.  The latter
require a further renormalization by the proper wave function renormalization constants, 
which at the level of the effective potential
is absorbed in a renormalization of $\chi^2_b \rightarrow Z_{\Phi} \chi^2_R$. The 
wave-function renormalizations must be calculated differently from the effective 
potential, since their calculation requires
one-loop contributions at non-zero momentum.  
\item After restricting the Hamiltonian to the gauge invariant subspace, one can
pass on to the Lagrangian density in terms of gauge invariant variables,
and write it in terms of the renormalized couplings, masses and fields
plus the counterterm Lagrangian. The counterterms are then required to
cancel the divergences in the proper 1PI Green's functions. It is at this
stage that we use the path integral representation discussed previously
to compute the wave function renormalization constants from the
one loop self-energies and vertex corrections. Such a calculation, although
straightforward,  involves non-covariant
loop integrals which are performed with the help of the appendix in
reference\cite{coulomb}. 
\end{enumerate}
For the renormalization of the effective 
potential only scalar wave function renormalization is needed beyond
the cancellation of the terms proportional to $\Delta_{\epsilon}$ in
(\ref{veffinv}). A lengthy but straightforward calculation of the relevant
one loop diagrams provides the renormalization
constants listed in Appendix B.

In terms of the renormalized couplings and expectation value of the
gauge invariant scalar field, the gauge invariant effective potential
in the $\msb$ renormalization scheme is given by
(here we drop the subscripts (R) in the renormalized quantities to
avoid cluttering of notation, but all quantities below are renormalized)

\begin{eqnarray}
\vef(\chi)  & = &   V_{cl}(\chi)+  \frac{1}{(4\pi)^2}\left\{
\frac{M^4_T[\chi]}{2}\left(\ln\left[\frac{M^2_T[\chi]}{\kappa^2}\right]-\frac{1}{2} \right)+  
 \frac{M^4_H[\chi]}{4}\left(\ln\left[\frac{M^2_H[\chi]}{\kappa^2}\right]-\frac{3}{2} \right)+ \right. \nonumber \\
& & \left. \frac{1}{4} \left(M^2_T[\chi]-M^2_g[\chi]\right)^2
\left(\ln\left[\frac{(M_T[\chi]+M_g[\chi])^2}{\kappa^2}\right]-\frac{3}{2} \right) \right. \nonumber \\
 &  & \left. 
-\frac{1}{2}M_T[\chi]M_g[\chi](M_T[\chi]-M_g[\chi])^2- M^4_F[\chi]\left(\ln\left[\frac{M^2_F[\chi]}{\kappa^2}\right]-\frac{3}{2} \right)
\right\} \label{veffinvren}
\end{eqnarray}

\section{Comparison with gauge-fixed results:}

A comparison with the effective potential in covariant $R_{\xi,u}$ 
gauges is established by adding a gauge fixing and Faddeev-Popov terms
to the Lagrangian density eq.(\ref{higgslag}):
\begin{eqnarray}
&& {\cal{L}}\rightarrow {\cal{L}}+{\cal{L}}_{gf}+{\cal{L}}_{FP} \label{totallag} \\
&& {\cal{L}}_{gf} = -\frac{1}{2\xi}(\partial_{\mu}A^{\mu}+\xi u e
\phi_2)^2 \label{gaugefix} \\
&& {\cal{L}}_{FP}= - \bar{c}{\partial^2}c-\xi u e^2 \bar{c} c
\phi_1 \label{faddeevpopov}
\end{eqnarray}

Special cases of this covariant gauge fixing procedure
include the  Landau gauges $ \left(\xi=0 \right)$,
 't Hooft $R_{\xi}$ gauges \cite{FLS}
($u=\left\langle \phi_1 \right\rangle_0$, the tree-level Higgs $vev$) and Fermi gauges $ \left( u=0 \right)$ 
where we have chosen the symmetry breaking expectation value along the
$\phi_1$ direction for convenience. At this stage we can just quote
the results of reference\cite{willray} adapted to the case treated 
here of an axial vector coupling with $N_f=1$, for arbitrary number
of flavors, the one-loop contribution from the fermion will be multiplied
by $N_f$. With $\varphi= \langle \phi_1 \rangle$  the expectation
value obtained in the gauge-fixed path integral, we obtain
the one loop effective potential in the $\msb$ scheme after renormalization of $\mu^2$, couplings and
wavefunctions\cite{willray}:

\begin{eqnarray}
\tilde{V}_{eff}\left[ \varphi; \xi; u \right] & = & {1 \over 2} \mu^2 \varphi^2 +
 {{\lambda} \over 4} \varphi^4 
+ {1 \over 4} {{H^4\left[ \varphi \right]} \over { \left( 4 \pi \right)^2}} 
\left[ \ln{{{H^2\left[ \varphi \right]} \over {\kappa^2}}} - {3 \over 2} \right]\nonumber \\
& &+ {3 \over 4} {{B^4\left[ \varphi \right]} \over { \left( 4 \pi \right)^2}} 
\left[ \ln{{{B^2\left[ \varphi \right]} \over {\kappa^2}}} - {5 \over 6} \right]
- {2 \over 4} {{G^4\left[ \varphi \right]} \over { \left( 4 \pi \right)^2}} 
\left[ \ln{{{G^2\left[ \varphi \right]} \over {\kappa^2}}} - {3 \over 2} \right]\nonumber \\
& &+ {1 \over 4} {{k_+^4\left[ \varphi \right]} \over { \left( 4 \pi \right)^2}} 
\left[ \ln{{{k_+^2\left[ \varphi \right]} \over {\kappa^2}}} - {3 \over 2} \right]
+ {1 \over 4} {{k_-^4\left[ \varphi \right]} \over { \left( 4 \pi \right)^2}} 
\left[ \ln{{{k_-^2\left[ \varphi \right]} \over {\kappa^2}}} - {3 \over 2} \right] \nonumber \\
& &-  {{Y^4\left[ \varphi \right]} \over { \left( 4 \pi \right)^2}} 
\left[ \ln{{{Y^2\left[ \varphi \right]} \over {\kappa^2}}} - {3 \over 2} \right]
+\xi e^2 u \varphi \left( \mu^2 + \lambda \varphi^2 \right) 
\left[\frac{-\Delta_{\epsilon}}{(4\pi)^2}\right]
\label{vef}
\end{eqnarray}
where 
\begin{eqnarray}
H^2\left[ \varphi \right] &= & \mu^2 + 3 \lambda \varphi^2\\
B^2\left[ \varphi \right]  &= & e^2 \varphi^2\\
G^2\left[ \varphi \right]  &= & \xi e^2 u \varphi \\
k_\pm^2\left[ \varphi \right]   & =& {1 \over 2} \left[ \mu^2 + \lambda 
\varphi^2 
+ 2 \xi e^2 u \varphi \right]  \pm \nonumber \\
& & {1 \over 2} \sqrt{\left( \mu^2 + \lambda \varphi^2  \right)
\left( \mu^2 + \lambda \varphi^2 + 4 \xi e^2 \varphi \left( u-\varphi \right)  \right)}\\
Y^2\left[ \varphi \right]   & = & y^2 \varphi^2
\end{eqnarray}
$H$, $B$, $Y$ and $G$ denote contributions from  Higgs, vector boson, 
 fermion and Fadeev-Popov ghost loops, respectively.   

Although the ``effective masses'' $H[\varphi] \; ; \; B[\varphi]\; ; \; 
Y[\varphi]$
are {\em formally} similar to the Higgs, transverse and fermion
 effective masses
given in equations (\ref{fermionfreq},\ref{transfreq},\ref{realfreq}) upon the 
replacement $\varphi \rightarrow |\chi|$, we want to emphasize that
$\varphi_{\rm ex}$ is the expectation value of a gauge dependent field in a gauge-fixed state, 
whereas $\chi$ is a true gauge invariant order
parameter, the expectation value of a gauge invariant field in a gauge
invariant state. The UV pole in the gauge-fixed effective potential
(\ref{vef}) which is not removed by couplings and wave-function renormalizations 
has been discussed in detail in\cite{willray}. It can
be removed by a shift in $\varphi$ consistently in the loop expansion and
is a consequence of the particular choice of $R_{\xi,u}$ gauge-fixing that
breaks the global $U(1)$ symmetry explicitly. 

Even for the Landau gauge, corresponding to the choice $\xi=0$, leading
to $k_+^2=\mu^2 + \lambda \varphi^2 ; \; k_-^2=0$ the difference of the gauge sector
contributions to the gauge fixed and the gauge invariant effective potential is clear.

The first {\em physical} quantity that we must compare is the value
of both the gauge invariant and the gauge fixed effective potentials
at their respective minima. The minima are obtained from the 
conditions
\begin{equation}
\left[ {{\partial \vef \left[  |\chi| \right] } \over {\partial |\chi|}} \right] \; \; ; \; \; 
\left[ {{\partial \tilde{V}_{eff} \left[  \varphi; \xi;u \right] } \over {\partial \varphi}} \right]=0
\end{equation}

Writing the stationary values of $|\chi| \; ; \; \varphi$ in a formal 
loop expansion as\cite{willray}
\begin{equation}
|\chi|_{\rm ext}= |\chi|_0 + |\chi|_1 + \cdots \; ; \; 
\varphi_{\rm ext} = \varphi_0 + \varphi_1 + \cdots
\end{equation}
we find the following relations that fix the minima to one-loop
order
\begin{eqnarray}
0&=& \left[ {{\partial V^0_{eff} \left[  |\chi| \right] } \over {\partial |\chi|}} \right]_{|\chi|_0} \label{tree1} \\
 |\chi|_1 &=& - \left[ {{\partial V^1_{eff} \left[  |\chi| \right] } \over {\partial |\chi|}} \right]_{|\chi|_0}
\left\{\left[ {{\partial^2 V^0_{eff} \left[  |\chi| \right] } \over {\partial |\chi|^2}} \right]_{|\chi|_0}\right\}^{-1}\label{onelup1} \\
0  &=& \left[ {{\partial \tilde{V}^0_{eff} \left[ \varphi;\xi ;u \right] } \over {\partial \varphi}} \right]_{\varphi_0}  \label{tree2} \\
 \varphi_1(\xi;u) &=& - \left[ {{\partial \tilde{V}^1_{eff} \left[ \varphi;\xi;u \right] } \over {\partial \varphi}} \right]_{\varphi_0}
\left\{\left[ {{\partial^2 \tilde{V}^0_{eff} \left[\varphi;\xi;u \right] } \over {\partial \varphi^2}} \right]_{\varphi_0}\right\}^{-1}\label{onelup2}
\end{eqnarray}
the solutions of eq.(\ref{tree1},\ref{tree2}) are obviously the
tree level vacuum expectation values $|\chi|_0 = \varphi_0 = \sqrt{-\mu^2/\lambda}$ but the solution of eq.(\ref{onelup2}) is
dependent on the gauge fixing parameters $\xi, u$\cite{willray}.
 However, upon 
inserting the solutions of the minima equation consistently up to one
loop in the respective expressions for the effective potentials, we
find that the value of the minima for the gauge-fixed and
the gauge invariant effective potential {\em are the same}. This is
an important result: whereas the gauge invariant effective potential
as constructed has the meaning of an energy of constrained physical
states, no such interpretation is available for the gauge-fixed result. 
However at the minimum, we see that the gauge-fixed effective potential
agrees with the gauge invariant result and therefore the values of the
extrema provide gauge invariant information on the energetics of constrained states. 

The second derivative of the gauge fixed effective potential at the
 minimum is seen to be a gauge dependent quantity. Although the pole
mass of the excitations are on-shell quantities and must therefore
be gauge invariant, the second derivative of the effective potential 
corresponds to the value of the  one-particle Green's function at 
zero four momentum transfer.

\begin{eqnarray}
\left[ {\partial^2 \vef } \over {\partial \varphi^2} \right]_{\varphi=\varphi_0}
&=&
2 \lambda \varphi_0^2 +
{{\varphi_0^2} \over {16 \pi^2}} \Biggl\{
6 \lambda^2 \left[ 4 \ln{{2 \lambda \varphi_0^2} \over {\kappa^2}} - 1\right] 
+ 9 e^4 \ln{{e^2 \varphi_0^2} \over {\kappa^2}} + \nonumber \\
& & \left[ 2 \lambda^2 + {1 \over 2} \lambda\xi e^2 + \left( \xi e^2 \right)^2 \right]
\ln{{\xi e^2 \varphi_0^2 } \over {\kappa^2}} - \lambda \xi e^2 
 - 16 N_f y_t^4 \left[ 3 \ln{{y_t^2 \varphi_0^2} \over {\kappa^2}} - 1 \right]
\Biggr\}
\end{eqnarray}

We note that this expression is divergent in the Landau gauge ($\xi=0$).
Direct calculation in Landau gauge shows that the divergence is due to the 
Goldstone boson term in the effective potential.

\section{RG improvement and the Higgs mass bound}

The vacuum stability bound on the Higgs mass arises from imposing the 
requirement that the electroweak vacuum be the global minimum of the effective potential.  
In fact, the Standard Model is thought to be the low-energy 
effective theory of some more fundamental high-scale theory.  Thus it is only
consistent to demand that the electroweak vacuum is the absolute minimum
up to the scale at which the effects of the `new physics' (those not incorporated 
into the low-energy effective theory) become significant.  In the context of 
the effective potential we associate this scale with a value of the
elementary scalar field and hence insist that the electroweak minimum is the 
global minimum of the effective potential up to that value of the field (however, see \cite{HungSher}).  For the
elementary scalar field effective potential, however, the value that the effective
potential assumes at some expectation value of the scalar field is explicitly gauge dependent.  Thus, 
the statement that the value of the effective potential exceeds that of 
the electroweak minimum up to some scale $\Lambda$ is also gauge dependent.
This gauge dependence in turn infects the lower bound on the Higgs mass
derived from that ans\"atze. 

The PEP does not suffer from this gauge ambiguity.  The condition that
the electroweak minimum is the global minimum up to some maximum value
of the gauge invariant order parameter provides a gauge invariant 
means of defining the vacuum instability scale and through that a gauge
invariant lower bound on the Higgs mass.

Using the PEP, we demonstrate how,
to derive the lower bound on the Higgs mass 
in the toy model, which displays the same qualitative features as the Standard Model with respect to 
vacuum stability.  Unlike the conventional approach using
the elementary field effective potential, the bound obtained from the PEP formalism
will be manifestly gauge invariant.  The extension of the approach to 
nonabelian gauge theories such as the Standard Model
requires the extension of the PEP formalism
to that context, we see no serious obstacles to such a program.

It is known (in the context of the elementary field effective potential) 
that the simple perturbative effective potential is often inadequate
for the study of the theory at large field values  
due to large logarithms which ruin the convergence of the perturbative loop 
expansion\cite{ColemanWeinberg,sher}.  The range of validity of the 
perturbative expansion can be enhanced, however, through the
 use of the renormalization 
group.   
The invariance of the full (all-orders) effective potential under a change in 
$\msb$ renormalization scale can be written as a first-order differential equation
\begin{equation}
\kappa{d \over {d \kappa}}{\hat{V}_{\rm eff}}\left( \varphi,\mu,\lambda,e,y_t,\xi,\kappa \right)=0
\end{equation}
where $\xi$ denotes a generic gauge parameter, which always arises in defining the gauge-fixed elementary
field effective potential.
This equation may be solved using familiar methods in terms of the 
$\beta$ and $\gamma$ functions.  These in turn may be calculated in a loop 
expansion from the $\msb$ counterterms of the theory.  Whereas the 
unimproved effective potential was reliable only for field values at which
\begin{equation}
{\hat{g}^2 \over 16 \pi^2} \ln{{\hat{g}^2 \varphi^2} \over {\kappa^2}} \ll 1
\end{equation}
the solution to the RG equation will be reliable as long as the running couplings
(generically $\hat{g}^2(s)$) are small.  

Similar considerations are applicable to the PEP (but now without the gauge parameter $\xi$).    The PEP is also
independent of changes in renormalization scale. The PEP has been identified as a matrix
element of the physical Hamiltonian in (physical) states. The Hamiltonian is part of the
energy momentum tensor which is a conserved current. Conserved currents do not acquire
anomalous dimensions and are finite after field independent subtractions (normal ordering
in the free field vacuum) in terms of the renormalized parameters and fields. In dimensional
regularization these normal ordering divergences vanish identically. Alternatively, that
the PEP is renormalization group invariant can also be seen in the same manner as for the
elementary field effective potential, by returning to the path integral,
expressing the PEP as a  sum of $\Phi^n \Gamma^{(n)}_\Phi$ 
at zero external momentum via the momentum expansion of the corresponding
effective action.  The multiplicative renormalization factors cancel, and since the renormalized 
effective potential equal to the bare effective potential it is also $\kappa $ -independent.
The analogous RG equation may then be written:
\bdm
\kappa{d \over d \kappa}{\vef}\left( \chi,\mu,\hat{g},\kappa \right)=\left[ 
\kappa{ \partial \over \partial \kappa} + \beta_{\hat{g}} { \partial \over \partial \hat{g}} -  \gamma_\mu \mu{\partial \over \partial \mu}  - \gamma_\Phi \chi {\partial \over 
\partial \chi} \right]
 \vef \left( \chi,\mu,\hat{g},\kappa \right)=0
\edm
where 
\bea
\beta_{\hat{g}}&=&\kappa {d \hat{g} \over d\kappa}  \nonumber \\
\gamma_{\Phi}&=&-{\kappa \over \Phi}{ d \Phi \over d \kappa}=
{1 \over 2} {\kappa \over Z_{\Phi}} {d Z_{\Phi} \over d \kappa} \nonumber \\
\gamma_{\mu}&=&-{\kappa \over \mu}{ d \mu \over d \kappa}=
{1 \over 2} {\kappa \over Z_{\mu^2}} {d Z_{\mu^2} \over d \kappa}
\eea
or, writing $\Phi=s \Phi_i$ (where $\Phi_i$ is some arbitrary initial scale, for example the electroweak scale)
\bdm
\left[ \kappa{\partial \over \partial \kappa} + \beta_{\hat{g}} {\partial \over \partial \hat{g}} -  \gamma_\mu \mu{\partial \over \partial \mu}  - \gamma_\Phi s {\partial \over \partial s} \right] \vef 
\left( s \chi_i,\mu,\hat{g},\kappa \right)=0 \label{RG1}
\edm
Here $\hat{g}$ represents the set of couplings $\lambda$, $g$, and $y_t$.
Using dimensional analysis we can also write
\bdm
\left[ \kappa {\partial  \over \partial \kappa} + \mu {\partial \over \partial \mu} + s  {\partial \over \partial s} - d\right] \vef\left( s \chi_i,\mu,\hat{g},\kappa \right)=0.
\label{dimanal}
\edm
Combining eq.\ref{RG1} and eq.\ref{dimanal} gives the RG equation
\bdm
\left[ \mu (\gamma_\mu - 1){\partial \over \partial \mu} + \beta_{\hat{g}} {\partial \over \partial \hat{g}} 
- (\gamma_\Phi + 1) s  {\partial \over \partial s} + d\right] \vef\left( s \chi_i,\mu,\hat{g},\kappa \right)=0.
\edm
It can then be shown that solution of the $\msb$ 
RG equation for the effective potential is 
\begin{equation}
{\vef\left( s \chi_i,\hat{g}_i,\mu_i,\kappa \right)}=
\left[\zeta(s)\right]^d\vef\left(\chi_i, \hat{g}\left( s, \hat{g}_i \right),\mu \left( s, \mu_i \right), \kappa \right) \label{RGsol}
\end{equation}  
where 
\bdm
\zeta(s)=\exp{\left[\int_{0}^{\log{s}}{1 \over \gamma_{\Phi}\left( x \right) + 1}dx\right]}, 
\edm
 $\hat{g}(s,\hat{g}_i)$ is the solution to the equation
\bea
s {{d \hat{g}(s) } \over {d s}}&=&\bar{\beta}_g\left( \hat{g}\left( s \right) \right)={\beta_g \left( \hat{g} \left( s \right)\right)\over 1 + \gamma_\Phi\left( \hat{g}\left( s \right) \right)}  \nonumber \\
\hat{g}(0)&=&\hat{g}_i
\eea
and 
\bdm
\mu(s,\mu_i)=\mu_i \exp{\left[\int_0^{\log{s}} \left[  {\gamma_{\mu}(x) - 1} \over {\gamma_{\Phi}(x) + 1} \right] dx\right]}
\edm
It is convenient to separate the tree-level $\vef$ and its one-loop corrections 
by writing eq.\ref{RGsol} in the form
\bdm
{\vef\left( s \chi_i,\hat{g}_i,\mu_i,\kappa \right)}=
\left[\zeta(s)\right]^d \left[ {1 \over 2} \left( m^2(s) + \Delta m^2( \chi_i)\right) \chi_i^2 + {1 \over 4} \lef(s,\chi_i) \chi_i^4\right]
\edm
\bdm
\lef(s,\chi_i)= \lambda\left( s \right) + \Delta \lambda\left( \chi_i \right)
\edm
where $ \Delta m^2$ and $ \Delta\lambda$ contain the loop corrections to $\vef$.

The $\beta$ and $\gamma$ functions and $\vef$
can then be calculated to the desired loop order.  
It has been shown that using the n-loop 
effective potential and the $n+1$ loop $\beta$ functions sums up all logs of the
form $\log{s}, \log^2{s}, \ldots, \log^n{s}$ \cite{Bando,Kastening_RG}.  
The one-loop $\msb$ $\beta$ functions for our model are given
explicitly in Appendix B.
$\Delta\lambda$ can be extracted directly from eq.(\ref{veffinv}) and to
one-loop order is
\bea
\Delta\lambda(s,\chi_i)  &=&  \frac{4}{(4\pi)^2 \chi_i^4} \Bigg\{
 \left. \frac{M^4_T[\chi_i]}{2} \left(\ln\left[\frac{M^2_T[\chi_i]}{\kappa^2}\right]-\frac{1}{2} \right)+ \frac{M^4_H[\chi_i]}{4}\left(\ln\left[\frac{M^2_H[\chi_i]}{\kappa^2}\right]-\frac{3}{2} \right)+ \right. \nonumber \\
& & \left. \frac{1}{4} \left(M^2_T[\chi_i]-M^2_g[\chi_i]\right)^2
\left(\ln\left[\frac{(M_T[\chi_i]+M_g[\chi_i])^2}{\kappa^2}\right]-\frac{3}{2} \right)
\right. \nonumber \\
& &
-\frac{1}{2}M_T[\chi_i]M_g[\chi_i](M_T[\chi_i]-M_g[\chi_i])^2-  \left. M^4_F[\chi_i]\left(\ln\left[\frac{M^2_F[\chi_i]}{\kappa^2}\right]-\frac{3}{2} \right) \right.
\Bigg\}
\eea  

In the Standard Model the top quark term dominates $\beta_{\lambda}$.
In this toy model the fermion plays the same role.
It is the large negative fermion loop contribution that drives 
 $\lambda(s)$ negative at large 
$s$.  At the leading-log level,
if the tree-level RG-improved effective potential is run
with the one-loop $\beta$ function, at large fields it is well-approximated by
neglecting the term quadratic in $\chi$ and is given by $\vef[s,\chi_i]= {1 \over 4} \lambda(s) [\chi_i \zeta(s)]^4$.  Very near the value of $s$ at which $\lambda(s)$ turns negative, $\vef$ falls below the electroweak
minimum.  We define the scale $s$ at which $\vef(s) = \vef(\sew)$ to be the vacuum instability scale, $\svi$.
If the electroweak minimum is to remain the global minimum of
the effective potential, the theory must somehow be incomplete and new
high-scale physics must contribute in some important way before the vacuum
instability scale.  

To extract a lower bound on the Higgs mass, we choose a vacuum instability 
scale $\svi$ where by definition $\vef(\svi)=\vef(\sew)$.  In the approximation
that $s \gg 1$, the contribution to $\vef(\svi)$ from the $m^2$ terms is small, 
and $\vef (\svi) \approx  {1 \over 4} \lef(\svi) \chi_i^4 \left[ \zeta(\svi) \right]^4 $.  Further, 
for $s \gg 1$, the value of $s$ at which $\vef(s)=0$ is very close to that at 
which it equals $\vef(\sew)$.  Thus it is numerically a good approximation
to take \cite{CEQ1,CEQ2,CEQ3}
\bdm
\vef(\svi,\chi_i) \approx{1 \over 4} \lef(\svi) \chi_i^4 \left[ \zeta(\svi) \right]^4 \approx 0.  
\label{vefzero}
\edm
Since $\zeta(s)$ is positive, eq.\ref{vefzero} implies
\bdm
\lef(\svi)=\lambda(\svi)+\Delta\lambda(\svi)=0 \Longrightarrow \lambda(\svi)=-\Delta\lambda(\svi) 
\edm
Thus the constraint that the electroweak minimum be the absolute minimum of the effective 
potential up to some high scale translates to a high-scale boundary condition on $\lambda(s)$.  Run down to
the low scale, the result is
 a value of $\lambda_i$ below which the electroweak vacuum
is unstable.  This can in turn be converted to a lower bound on the Higgs pole mass.  

This is completely analogous to the procedure with the elementary field effective potential.  
In that case, however, the expression for $\Delta\lambda$ is explicitly
dependent on the gauge parameters, which results in a gauge dependence in 
the Higgs mass bound \cite{willray}.  
In the current formulation, $\Delta\lambda$ is by 
construction gauge-invariant.  The steps all follow through in exactly the same fashion, but the Higgs mass
bound obtained in the end is gauge invariant.

\section{Numerical Results}

Here we demonstrate the formalism developed in the previous section by 
applying it to obtain numerical values for $\lambda_i$, assuming values for the other couplings at 
the electroweak scale and assuming some value for $\svi$.
Since this is only an abelian toy model, we cannot draw conclusions about 
the Standard Model.  Instead, we compare the results of the gauge 
invariant formulation with the elementary field effective potential formulation
within this model for several different sets of parameters.

For illustrative purposes it will suffice to use only
one-loop $\beta$ functions and the one-loop corrections to the effective 
potential, which will sum the leading logs.  Furthermore, at the leading log level
it is consistent to run the couplings in the tree-level part of the effective potential 
only and to leave the couplings in the one-loop terms fixed at their initial values.
The additional effort required to reduce the $\kappa$ dependence of the
results and improve numerical accuracy  by calculating higher loop effects 
is not justified for a toy model.

In Figure 1 we plot the log of the vacuum instability scale $\svi$ 
as a function of $\lambda_i$ and $e^2_i$ for $y_{t,i}^2=0.5$ and $\kappa=v$, as 
calculated using the PEP.  We have also performed a similar calculation using the
Landau gauge elementary field effective potential (setting $\xi =0$ in \ref{vef})
 and found the results to differ only by a few percent.  This is shown explicitly in Figure 2.
Here the difference 
between the $\log{\svi}$ calculated using the different effective potentials is plotted,
and the difference is shown to be very small. Thus, in this model 
 the Landau gauge elementary field effective potential treatment and a PEP treatment give 
very similar numerical results at one loop level.
 
We note that the similarities of the numerical results between the PEP and the Landau
gauge fixed effective potential are not obvious {\em a priori}. In a gauge fixed formulation
there is  {\em a priori} no reason to expect the results in one gauge to be numerically
 superior to those in another gauge.

The similarities between the numerical results on the scale of new physics and therefore
on the bound on the Higgs mass between the gauge invariant and gauge fixed
formulation is due to the fact that the gauge couplings considered are relatively weak. 
 For stronger coupling the gauge and scalar sectors will be more important in 
$\Delta\lambda$ and the differences more noticeable.  However, in that case a
one-loop calculation will likely not be adequate and will thus be beyond the
scope of our simple one-loop analysis.  

\section{Conclusions}

The usual formulation of bounds on the Higgs mass from 
the RG improved elementary field effective potential is afflicted with gauge 
dependence that renders it unsatisfactory.  This gauge dependence is
the result of the gauge dependence of the elementary field effective potential 
itself, and it suggests that the elementary field effective potential is not the
appropriate tool for the analysis. The error stemming from the gauge variance in
calculations in a specific gauge-fixing scheme cannot be inferred because the results can
vary in a wide range by varying the gauge parameter.

 We have presented an alternative 
approach, using the recently-introduced PEP \cite{boyan}, to calculate
the Higgs mass bound in a simplified abelian  model which displays the same 
vacuum stability problem as the Standard Model. We have explicitly provided a
gauge invariant construction of the effective potential in terms of a gauge invariant
order parameter that serves as a signal of symmetry breaking. This effective potential 
is unambiguously identified with the physical energy of a configuration and therefore
provides reliable estimates from vacuum stability analysis. We constructed this 
gauge invariant effective potential to one loop order and improved it via RG. 

  Compared to the analogous calculation in the Landau gauge the numerical results are 
similar when the gauge couplings are weak.  While this suggests that Landau gauge
calculations in the Standard Model may  give numerically similar
results to a gauge-invariant treatment, it does not provide justification for the principle of using the gauge-dependent effective potential to place bounds on gauge invariant quantities. 

Although not guaranteed {\em a priori}, our result on the numerical similarities between
the scales and bound obtained from PEP and the Landau gauge-fixed effective potential 
 suggest that in the case of weak gauge couplings, the Landau gauge effective potential
provides a qualitatively reliable estimate. However, the PEP is necessary to establish
the ``error'' estimate. In that respect, our results provide a tentative credibility to
bounds based on Landau gauge effective potential calculations.  

An extension of the PEP formalism to nonabelian gauge theories would permit a gauge invariant calculation of the Higgs mass bound in the Standard Model using the PEP.  Such a gauge invariant calculation is required to provide justified error
bounds on the Higgs mass from vacuum stability considerations.  We hope to provide such an extension in the near future.

\acknowledgements

D.Boyanovsky acknowledges support from the N.S.F. through grant No: PHY-9302534. We wish to thank Paresh Malde for useful correspondence
and Tony Duncan for useful conversations.

\appendix
\section{Dimensionally-Regularized Integrals}

Most of the integrals that arise in the calculation of the loop corrections to
the gauge-invariant effective potential are familiar.  The contributions of 
eqns. \ref{transfreq} - \ref{imfreq} generate integrals of generic form
\begin{equation}
  \mu^\epsilon \int{ {{d^{d-1}k}  \over {\left( 2 \pi \right)^{d-1}} }
\; \sqrt{k^2 + m^2} }= {{m^4} \over {32 \pi^2}} \left[ -\Delta_\epsilon + \ln{{m^2} \over{\mu^2}} - {3 \over 2} \right]
\end{equation}
The more complicated term involving eq. \ref{newmomcoord}, however, requires additional effort to extract the ${\cal O}(\epsilon^0)$ piece.  
The angular integration gives
\begin{eqnarray}
 \mu^\epsilon \int{ {{d^{d-1}k}  \over {\left( 2 \pi \right)^{d-1}} }
\; \sqrt{ {{(k^2 + M^2_T) (k^2 + M^2_g)} \over {k^2}}} }&=& 
{ \mu^\epsilon \Omega_{d-2} \over {\left(2 \pi \right)^{d-1}}} 
\int_{0}^{\infty}{dk \; k^{d-2} \sqrt{ {{(k^2 + M^2_T) (k^2 + M^2_g)} \over {k^2}}}  } \\
&=&{1 \over 2} { \mu^\epsilon \Omega_{d-2} \over {\left(2 \pi \right)^{d-1}}}  
M_T^d  \int_{0}^{\infty}{du \; u^{d-4 \over 2} \sqrt{1+u } \sqrt{r+u}}
\end{eqnarray}
where $
\Omega_{d}= 2^d \pi^{d \over 2} {{\Gamma\left[ {d \over 2}  \right]} \over  {\Gamma\left[  d \right]}}$ and $r={M_G^2 \over M_T^2}$.
Using eq.3.197.1 and eq.9.111.2 of  \cite{GR} this can be written
\begin{eqnarray}
M_T^d{ \mu^\epsilon \Omega_{d-2} \over {\left(2 \pi \right)^{d-1} \Gamma\left[- {1 \over 2}  \right]}}  & & \Bigg\{ \Gamma\left[ -{d \over 2} \right] \Gamma\left[ {  d-1 \over 2} \right]  \; {_2F_1}\left[ -{1 \over 2}, -{d \over 2}, { 3-d  \over  2}, r \right]  \nonumber \\
& +&   r^{d-1 \over 2}
\Gamma\left[ {{d-2} \over 2} \right] \Gamma\left[ -{{d-1} \over 2}\right]{_2F_1}
\left[ -{1 \over 2}, {{d-2} \over 2}, {{d+1} \over 2}  , r \right] \Bigg\}
\end{eqnarray}
where $d=4 - \epsilon$.  The second term 
is finite for $\epsilon \rightarrow 0$ and so $\epsilon=0$ may be take immediately.
It may be written as
\begin{equation}
{M_T^4 \over 4 \pi^2} {1 \over 4} \left\{  -\sqrt{r} 
\left( 1+r \right) +\left( 1-r \right)^2 {\rm arctanh}{\sqrt{r}}   \right\}.
\end{equation}
The first term contains a ${1 \over \epsilon}$
pole arising from the $ \Gamma\left[ -{d \over 2} \right] $ factor.  A Laurent expansion 
about $\epsilon=0$ to ${\cal O}(\epsilon^0)$ is necessary, which requires the 
${\cal O}(\epsilon)$ terms of the hypergeometric function.  This is obtained 
by writing  $_2F_1\left[ -{1 \over 2} ,-2 + {\epsilon \over 2},{ -1 + \epsilon  \over 2},r\right]$ as a series expansion in $r$ in terms of Pochhammer symbols,
carrying out the expansion in $\epsilon$ on each term, and resumming the
resulting series.  To ${\cal O}(\epsilon)$:
\begin{eqnarray}
{_2F_1}\left[ -{1 \over 2} ,-2 + {\epsilon \over 2},{ -1 + \epsilon  \over 2},r\right]&=& {_2F_1}\left[ -{1 \over 2} ,-2 ,-{1 \over 2} ,r\right] + \epsilon
{\partial \over {\partial \epsilon}} \; {{_2F_1}\left[ -{1 \over 2} ,-2 + {\epsilon \over 2},{ -1 + \epsilon  \over 2}, r \right] }_{\epsilon=0} \nonumber \\
&=& \left( 1-r \right)^2 +\sum_{n=0}^{\infty}{
{ \epsilon {\partial \over \partial \epsilon} 
\left[
{
{\left( -{1 \over 2} \right)_n \left( -2 + {\epsilon \over 2} \right)_n } 
\over 
{\left({ -1 + \epsilon  \over 2} \right)_n} 
}
\right]_{\epsilon=0} {r^n \over n!} 
} }\nonumber \\
&=& \left( 1-r \right)^2 +\epsilon \sum_{n=0}^{\infty}{
{r^n \over n!} 
\left[  {\partial \over \partial \epsilon} \left( -2 + {\epsilon \over 2} \right) \right]_{\epsilon=0} } \nonumber \\
& & -\epsilon \sum_{n=0}^{\infty}{
{r^n \over n!} 
\left[  
 {\left( -2 \right)_n \over \left( -{1 \over 2} \right)_n} 
 {\partial \over \partial \epsilon} \left(  {-1+ \epsilon \over 2} \right)_n 
\right]_{\epsilon=0}
} 
\label{hypergeomexpn}
\end{eqnarray}
Upon carrying through the differentiation the first term of eq.\ref{hypergeomexpn} yields an infinite series that can be resummed.
The second series terminates due to the factor $\left(-2 \right)_n$.
The final result is:
\begin{equation}
{_2F_1}\left[ -{1 \over 2} ,-2 + {\epsilon \over 2},{ -1 + \epsilon  \over 2},r\right]= \left( 1-r \right)^2 - \epsilon \left[ 4 r + \left( 1-r \right)^2 \ln{\left[1-r\right]} \right]
\end{equation}
This is combined with the series expansions of the gamma functions
and other prefactors, and terms to ${\cal O}(\epsilon^0)$ are collected.
The result is: 
\begin{eqnarray}
\lefteqn{
 \mu^\epsilon \int{ {{d^{d-1}k}  \over {\left( 2 \pi \right)^{d-1}} }
\; \sqrt{ {{(k^2 + M^2_T) (k^2 + M^2_g)} \over {k^2}}} }=
{M_T^4 \over 32 \pi^2} \bigg\{  \left( 1-r \right)^2  \left( \Delta_\epsilon + {3 \over 2} - \ln{ M_T^2 \over \mu^2 }\right) 
} \nonumber \\
& & + 2 \left( - \sqrt{r}  \left( 1+r \right)+ \left( 1-r \right)^2 {\rm arctanh}\sqrt{r}  \right)
+ 4 r + \left( 1-r \right)^2 \ln{\left( 1-r \right)}  \bigg\} \nonumber \\
&=& -{\left( M_T-M_g \right)^2 \over 32 \pi^2} \left\{  \left( M_T + M_g \right)^2 \left( \Delta_\epsilon + {3 \over 2} - \ln{\left( M_T + M_g \right)^2 \over \mu^2 }  \right) 
+ 2 M_T M_g  \right\}
\end{eqnarray}

\section{$\msb$ Counterterms and $\beta$ Functions}

\subsection{$\msb$ Counterterms}
\begin{eqnarray}
\delta Z_{\Phi} & = &  (2e^2-2y^2) 
\left[ \frac{\Delta_{\epsilon}}{16\pi^2} \right] \label{Phiwavefunc} \\
\delta Z_{\mu^2}& = & (4\lambda -3e^2 +2y^2)
\left[ \frac{\Delta_{\epsilon}}{16\pi^2} \right] \label{massrenfunc} \\
\delta Z_{\lambda} & = & (10 \lambda -6e^2 + 3\frac{e^4}{\lambda}+
4y^2 -4 \frac{y^4}{\lambda})
\left[ \frac{\Delta_{\epsilon}}{16\pi^2} \right] \label{scalarcoupren} \\
\delta Z_{A} & = & -\frac{2e^2}{3} 
\left[ \frac{\Delta_{\epsilon}}{16\pi^2} \right] \label{gaugewavefunc} \\
\delta Z_{e} & = & \frac{e^2}{3} 
\left[ \frac{\Delta_{\epsilon}}{16\pi^2} \right] \label{gaugecoupren} \\
\delta Z_{y} & = & (-\frac{3e^2}{4} + 2 y^2) 
\left[ \frac{\Delta_{\epsilon}}{16\pi^2} \right] \label{yukawaren} 
\end{eqnarray}

\subsection{$\beta$ Functions}

\bea
{\beta}_{\lambda} &=&{1 \over {16 \pi^2}} \left( 20 \lambda^2 + 6 e^4 - 8 y^4 - 12 e^2 \lambda + 8 \lambda y^2 \right)\\
{\beta}_{e} &=& {1 \over {16 \pi^2}}  \left( {2 \over 3} e^3 \right)\\
{\beta}_{y} &=& {1 \over {16 \pi^2}} y \left(- {3 \over 2} e^2 + 4 y^2 \right) \\
{\gamma}_{\Phi} &=& {1 \over {16 \pi^2}} \left(-e^2 + y^2  \right) 
\eea

\begin{figure}
\centering
\epsfig{file=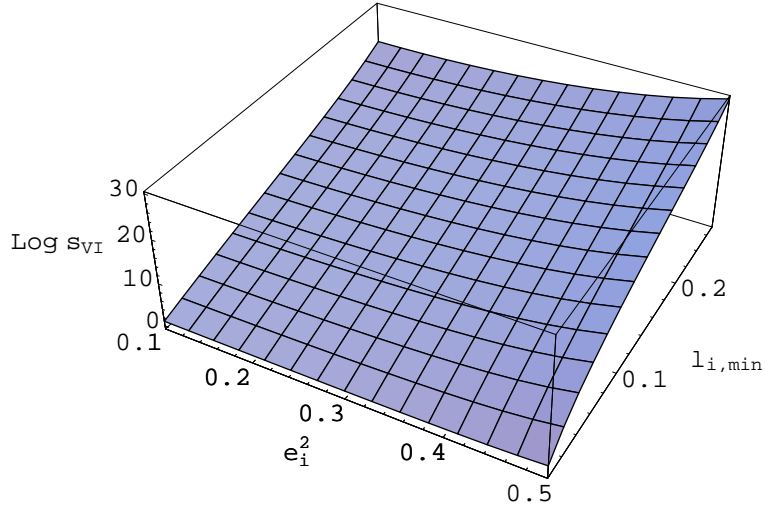,width=4in}
\caption{ $\log{\svi}$ for the PEP as a function of $e^2_i$ and $\lambda_{i,min}$ for $y_t^2=0.5$, $\kappa=v$.}
\end{figure}

\begin{figure}
\centering
\epsfig{file=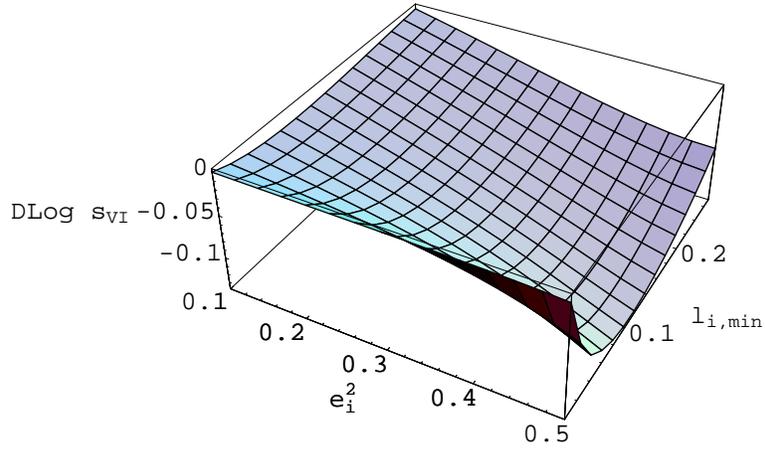,width=4in}
\caption{ The difference between $\log{\svi}$ calculated from the PEP and $\log{\svi}$ calculated from the Landau gauge elementary field effective potential, as a function of $e^2_i$ and $\lambda_{i,min}$, for $y_t^2=0.5$, $\kappa=v$.}
\end{figure}

\end{document}